\documentclass[twocolumn,showpacs,preprintnumbers,amsmath,amssymb]{revtex4}
\usepackage{graphicx}
\newcommand{\ket}[1]{|#1\rangle}
\newcommand{\bra}[1]{\langle#1|}

\begin{document}
\title{Phenomenological modelling for Time-Resolved Electron Paramagnetic Resonance in radical-triplet system}
\author{Wei Wu}\email{wei.wu@ucl.ac.uk}
\affiliation{UCL Department of Physics and Astronomy and London Centre for Nanotechnology, University College London, Gower Street, London WC1E 6BT, United Kingdom}

\begin{abstract}
The spin dynamics of radical-triplet system (RTS) has been calculated by using the Lindblad formalism within the theory of open quantum system. The single-radical-triplet system (SRTS) is considered here for single-qubit quantum gate operations while double-radical-triplet system (DRTS) for two-qubit operations. The environment effects taken into account include the spin-lattice relaxation of the triplet exciton and radical spin-$\frac{1}{2}$, the inter-system crossing process that induces the transition from singlet excited state to the triplet ground state, and the rather slow relaxation process from the triplet ground state back down to the singlet ground state. These calculations shown that the line shape broadening is strongly related to the exchange interaction between triplet and exciton, which can be understood as a spontaneous magnetic field created by the triplet renormalises the original spin-$\frac{1}{2}$ electron spin resonance spectra. This work will provide key information about the spin dynamics for building optically-controlled molecular quantum gate out of radical-bearing molecules. Moreover, this has generated the further theoretical question on how the mixture of fermion and boson behaves.
\end{abstract}


\maketitle
\section{Introduction}
Molecular-base electronics attracts much attention due to the advantages over conventional semiconductors \cite{sandrine07}, e.g., ease to engineer molecules, long spin-lattice relaxation time, and relatively less restrictive production conditions. In this sense, molecular-based electronics is a very promising way to tackle challenges obstructing the implementation of spintronics and quantum information processing (QIP). However, the manipulation of individual electron spin and controlling the interactions between electron spins in molecular species is difficult and crucial for the realization of spintronics and QIP in the molecular level. Time-Resolved Electron Paramagnetic Resonance (TR-EPR) \cite{ishii1998,corvaja2000,teki2000,teki2002,collison2004,franco2006} is a very useful tool for observing spin polarization and spin-spin interactions both for the molecules in the ground state and the excited state \cite{teki2002}.

Against the above background, the radical-triplet system (RTS) \cite{teki2002} which contains spin-bearing radicals and optical-active molecular species could be a prototype for molecular-based spintronics and QIP. From the literature \cite{collison2004}, RTS could contain either two radical spins or two radical spins and one photo-active coupler, i.e., single-radical-triplet system (SRTS) or double-radical-triplet system (DRTS). However, more radicals or photo-active species could be synthesized. TR-EPR is applied right after molecules are excited by laser shining from the ground state to the triplet excited state through inter-system crossing. The interactions between these spins and the spin polarization effect can be detected by TR-EPR. Compared with a large volume of experimental works on RTS there is very rare theoretical work to explain the observed EPR spectra from a point of view of the open quantum system, i.e., considering the linear response of an open quantum system which interacts with the environment when it is perturbed by an external field, rather than a closed quantum system although there is some theoretical work based on a rather ambiguous theory \cite{hudson1965,yamauchi2004}.

Therefore, in this paper we will first present a generalized model \ref{sec:1} for the quantum open system in the limit of Markov process, i.e., the system forgets its history much faster than its evolution. Following this general model, we will describe the time evolution of quantum states within RTS. And then we will produce some numerical results of TR-EPR simulations, and compare them with the experimental results. Furthermore we will base on our theoretical model to make some predictions which is not present in experiments, especially the spin correlations between two $\frac{1}{2}$-spins. At the end, we will draw some conclusions and have some discussions.

\section{Linear response theory for the quantum open system}\label{sec:1}
First let's start with a closed quantum system. Supposing we have a closed quantum system described by $\hat{H}_0$ and perturb this system by using an external field $\lambda f(t)\hat{B}$, where $\lambda$ is a small factor, the linear response of the closed system due to a time-dependent perturbation is described by the well known Kubo's formula \cite{kubo1954}, and the time-dependent and frequency-dependent response functions for an observable $\hat{A}$ read
\begin{eqnarray}
\phi_{\hat{A}\hat{B}}(\tau)&=&\frac{i}{\hbar}\theta(\tau)\langle[\hat{B}_I(0),\hat{A}_I(\tau)]\rangle
\\\chi_{\hat{A}\hat{B}}(\omega)&=&\int_{-\infty}^{+\infty}d\tau\phi_{\hat{A}\hat{B}}(\tau)\mathrm{exp}(i\omega\tau).
\end{eqnarray}
However, for an open quantum system we may have to access to a coarse-grained description in which the coherent Hamiltonian supplemented by incoherent terms representing the interaction with the environment. The most general description for the evolution of a quantum state of an open quantum system \cite{quantumopenbook} can be formulated by
\begin{eqnarray}
\frac{d\hat{\rho}}{dt}&=&-\frac{i}{\hbar}[\hat{H},\hat{\rho}]+\sum_\mu [\hat{A}_\mu\hat{\rho}\hat{A}_\mu^{\dagger}-\frac{1}{2}\{\rho,\hat{A}_\mu^{\dagger}\hat{A}_\mu\}]=\hat{\hat{\mathcal{L}}}\rho\nonumber
\\\rho(t)&=&e^{\hat{\hat{\mathcal{L}}}}\rho(0),
\end{eqnarray}\label{eqn:rhodynamics}
where the equation in the first line is so-called Lindblad equation.
Here $\hat{A}_\mu$ describes the quantum jumps due to the system-environment interactions and $\hat{\hat{\mathcal{L}}}$ is the Lindblad super operator. In order to derive formulae for the linear response of an open quantum system we should take some advantages of Heisenberg representation and Interaction representation. But before doing this, we can envisage that it is quite ambiguous to define the Hermitian conjugate of a super operator if we want to use the interaction representation. Therefore, first we define a scalar product of two operators by
\begin{eqnarray}
(\hat{A},\hat{B})=\mathrm{Tr}[\hat{A}^\dagger\hat{B}],
\end{eqnarray}
where $\dagger$ ensures the normal linearity conditions for a scalar product. This scalar product looks just like a vector scalar product on the 'extended vector' representations of the operators $\hat{A}$ and $\hat{B}$. With the help of this definition, we can now define a Hermitian conjugate of a super operator by stating that $\hat{\hat{\mathcal{L}}}$ should satisfy the following equations,
\begin{eqnarray}\label{eqn:spdefine}
(\hat{A},\hat{\hat{\mathcal{L}}}\hat{B})=(\hat{\hat{\mathcal{L}}}^\dagger\hat{A},\hat{B})=(\hat{B},\hat{\hat{\mathcal{L}}}^\dagger\hat{A})^*.
\end{eqnarray}

\subsection{Heisenberg representation}
With the definition stated in equation~\ref{eqn:spdefine} the Hermitian conjugate of a super operator corresponds to the ordinary matrix Hermitian conjugate of the four-index matrix representation as follows
\begin{equation}
(\hat{\hat{\mathcal{L}}}^\dagger)_{kl,ij}=(\hat{\hat{\mathcal{L}}}_{ij,kl})^*.
\end{equation}
In particular if we look at the expectation value of any Hermitian operator $\hat{O}$ with a time-independent Liuvillian
\begin{eqnarray}
\langle\hat{O}\rangle(t)&=& \mathrm{Tr}[\hat{O}\rho(t)]\nonumber
\\\nonumber&=&\mathrm{Tr}[\hat{O}e^{\hat{\hat{\mathcal{L}}}t}\rho(0)]
\\\nonumber&=&(\hat{O}^\dagger,e^{\hat{\hat{\mathcal{L}}}t}\rho(0))
\\\nonumber&=&(e^{\hat{\hat{\mathcal{L}}}^\dagger t}\hat{O}^\dagger,\rho(0))
\\&=&\mathrm{Tr}[\hat{O}(t)\rho(0)].
\end{eqnarray}
Therefore,
\begin{eqnarray}
\hat{O}(t)&=&e^{\hat{\hat{\mathcal{L}}}^\dagger t}\hat{O}
\\\partial_t\hat{O}(t)&=&\hat{\hat{\mathcal{L}}}^\dagger\hat{O}(t).
\end{eqnarray}
We can also generalize these to the case where the Liuvillian is time-dependent as follows
\begin{eqnarray}
\rho(t)&=&T\mathrm{exp}[\int_0^t\hat{\hat{\mathcal{L}}}(s)ds]\rho(0)
\\\hat{O}(t)&=&T\mathrm{exp}[\int_0^t\hat{\hat{\mathcal{L}}}^\dagger(s)ds]\hat{O}
\end{eqnarray}
\subsection{Interaction representation}
As shown in the following, we can see that starting from the interaction representation we are able to derive formulae for the linear response of an open quantum system systematically and easily. Let's first decompose the Liuvillian to two parts, i.e., $\hat{\hat{\mathcal{L}}}=\hat{\hat{\mathcal{L}}}_0+\hat{\hat{\mathcal{L}}}_1$, where $\hat{\hat{\mathcal{L}}}_1$ is small in some sense compared to $\hat{\hat{\mathcal{L}}}_0$. For the moment we suppose both of them are time-independent. Then we can define an interaction-picture operator by
\begin{equation}
\hat{O}_I(t)=e^{\hat{\hat{\mathcal{L}}}^\dagger_0 t}\hat{O},
\end{equation}
and then we can derive the density operator in the interaction picture
\begin{eqnarray}
\mathrm{Tr}[\hat{O}\rho(t)]&=&\mathrm{Tr}[\hat{O}_I(t)\rho_I(t)]\nonumber
\\\nonumber&&=(\hat{O},e^{\hat{\hat{\mathcal{L}}}_0 t}\rho_I(t))
\\&&=\mathrm{Tr}[\hat{O}\rho_I(t)].
\end{eqnarray}
Therefore we require
\begin{equation}
\rho_I(t)=e^{-\hat{\hat{\mathcal{L}}}_0 t}\rho(t)=e^{-\hat{\hat{\mathcal{L}}}_0 t}e^{\hat{\hat{\mathcal{L}}}t}\rho(0).
\end{equation}
The equation of motion for $\rho_I$ is now
\begin{eqnarray}\label{eqn:ipl}
\partial_t \rho_I(t)&=&\hat{\hat{\mathcal{L}}}_{1I}(t)\rho_I(t)
\\\hat{\hat{\mathcal{L}}}_{1I}(t)&=&e^{-\hat{\hat{\mathcal{L}}}_0t}\hat{\hat{\mathcal{L}}}_1e^{\hat{\hat{\mathcal{L}}}_0t}
\end{eqnarray}

\subsection{Linear response theory for stationary reference state}
We first derive the formulae in the case where $\rho_0$ is a stationary state, i.e., $e^{\hat{\hat{\mathcal{L}}}_0t}\rho_0=\rho_0$. Supposing $\hat{\hat{\mathcal{L}}}_1=\lambda f(t)\mathcal{B}$, where $\lambda$ is the small parameter, $\mathcal{B}$ is a general super operator, and $f(t)$ is an arbitrary function of time. Now we imagine the perturbation is turned on gradually from $t=-\infty$, and look for the change in the expectation value of an observable $\hat{A}$. We can evaluate this in the interaction representation
\begin{eqnarray}
\langle\hat{A}\rangle(t)-\langle\hat{A}\rangle_0&=&\lambda\int_0^\infty d\tau f(t-\tau)\mathrm{Tr}[\hat{A}_I(t)\mathcal{B}_I(t-\tau)\rho_0]\nonumber
\\&=&\lambda\int_{-\infty}^{\infty}f(t-\tau)\phi_{AB}(\tau)d\tau
\\\phi_{AB}(\tau)&=&\theta(\tau)\mathrm{Tr}[\hat{A}_I(\tau)\mathcal{B}\rho_0]=\theta(\tau)\mathrm{Tr}[\mathcal{B}^\dagger\hat{A}_I(\tau)\rho_0].
\end{eqnarray}
Notice that if the perturbation $\hat{\hat{\mathcal{B}}}$ is a Hamiltonian, so $\hat{\hat{\mathcal{B}}}=\frac{1}{i\hbar}[\hat{B},\cdot]$ and $\hat{\hat{\mathcal{B}}}^\dagger=\frac{i}{\hbar}[\hat{B},\cdot]=-\hat{\hat{\mathcal{B}}}$, then $\phi_{AB}(\tau)=\frac{i}{\hbar}\theta(\tau)\langle[\hat{B}_I(0),\hat{A}_I(\tau)]\rangle_0$, which is in agreement with Kubo's formulae. Now we can do the Fourier transformation $\chi_{AB}(\omega)=\int_{-\infty}^{\infty}d\tau\phi_{AB}(\tau)\mathrm{exp}(i\omega\tau)$. By using the equation $e^{\hat{\hat{\mathcal{L}}}_0\tau}=\sum_i\textbf{v}_R^{(i)}e^{\lambda_{0i}}\tau(\textbf{v}_L^{(i)})^T$, where $\lambda_{0i}$ are the eigenvalues of $\hat{\hat{\mathcal{L}}}_0$ and $\textbf{v}_{R,L}$ are the right and left eigenvectors of $\hat{\hat{\mathcal{L}}}_0$. After some algebra, finally we have
\begin{eqnarray}
\chi_{AB}(\omega+i\epsilon)&=&\frac{i}{\hbar}[(\hat{B}(\mathcal{L}_0^\dagger-i\omega-\epsilon)^{-1}\hat{A},\rho_0)\nonumber
\\\nonumber&&-((\mathcal{L}_0^\dagger-i\omega-\epsilon)^{-1}\hat{A}\hat{B},\rho_0)]
\\\nonumber&=&\frac{i}{\hbar}\mathrm{Tr}[\hat{A}(\mathcal{L}_0^\dagger-i\omega-\epsilon)^{-1}\hat{B}\rho_0
\\&&-\hat{B}\hat{A}(\mathcal{L}_0^\dagger-i\omega-\epsilon)^{-1}\rho_0]
\end{eqnarray}

\subsection{Linear response theory for time varying reference system}
If $\rho_0$ is not stationary of $\hat{\hat{\mathcal{L}}}_0$, we can still have the similar equation for the response function as follows
\begin{eqnarray}
\langle\hat{A}\rangle(t)-\langle\hat{A}\rangle_0&=&\lambda\int_0^\infty d\tau f(t-\tau)\phi(t-\tau)\nonumber
\\\nonumber\phi(t,\tau)&=&\theta(\tau)\mathrm{Tr}[\hat{A}_I(\tau)\mathcal{B}\rho_{0S}(t-\tau)]
\\&&\theta(\tau)\mathrm{Tr}[\hat{A}\mathcal{B}_I(-\tau)\rho_{0S}(t)],
\end{eqnarray}
where $\rho_{0S}(t)=e^{\mathcal{L}t}\rho_0$ is the unperturbed density operator in the \textsl{Schr\"{o}dinger picture} to time $t$. If we further assume that $f(t)=e^{\epsilon t}cos(\omega t)$. After some algebra similar to that in the case for stationary initial density matrix, we can have the following frequency-dependent response function
\begin{eqnarray}\label{eqn:chinonstationary}
\chi_{AB}(\omega+i\epsilon,t)&=&-\mathrm{Tr}[\hat{A}(\mathcal{L}_0^\dagger+i\omega-\epsilon)^{-1}\hat{\hat{\mathcal{B}}}\rho_{S0}(t)],
\end{eqnarray}
which is in agreement with the equation in ref.\cite{hudson1965}. This is also the basis for our simulation of TR-EPR spectra.

\section{System description and theoretical methodology}\label{sec:2}

In most of the TR-EPR experiments we are interested in molecules are exposed to the environments, in general a liquid solution. The spin-lattice relaxation time is very long ($\mu s$) at room temperature for the radical spins. And the system-environment interaction is almost through spin-orbital couplings and phonon effect. The inter-system crossing is really fast for some molecules due to the combination of spin-orbital coupling and electrical coupling between singlet excited state and triplet state. However the above two mechanisms lead to very slow relaxation back to the ground state; the life time of the phosphorescence is about $ms$ even at room temperature because the electrical coupling and spin-orbital coupling is relatively small. In summary, Markovian description which is appropriate for weak-coupling situations should be a good approximation for RTS. So the main task is to build up the effective spin Hamiltonian and quantum jumping operators, solve the resulting Liuville equation, and evaluate the spectra according to equation~\ref{eqn:chinonstationary}. However, before that we should also develop the formalism for the linear response of quantum open systems as TR-EPR experiments involve both system-environment interaction and the linear response of open systems.

\subsection{Effective spin Hamiltonian}
The effective spin Hamiltonian for SRTS reads
\begin{eqnarray}
\hat{H}_{\mathrm{SRTS}}^{spin}&=&g_{t}\mu_B\textbf{B}\cdot\hat{\textbf{S}}+g_{r}\mu_B\textbf{B}\cdot\hat{\textbf{s}}
+J\hat{\textbf{S}}\cdot\hat{\textbf{s}}\nonumber
        \\&&+D\hat{S}_z^2+E(\hat{S}_x^2-\hat{S}_y^2),
\end{eqnarray}
where the first is Zeeman energy; $g_t$ is the g-factor for the triplet and $g_r$ for radical, the second term is exchange interaction between radical and triplet; $J$ is the exchange constant, and zero-field splitting $D$ and $E$ for triplet. $\textbf{B}$ is magnetic field. $\textbf{S}$ is the triplet spin, and $\textbf{s}$ is $\frac{1}{2}$-spin. Similarly for DRTS the effective spin Hamiltonian is
\begin{eqnarray}
\hat{H}_{\mathrm{DRTS}}^{spin}&=&g_{t}\mu_B\textbf{B}\cdot\hat{\textbf{S}}+g_{r}\mu_B\textbf{B}\cdot(\hat{\textbf{s}}_1+\hat{\textbf{s}}_2)\nonumber
    +J\hat{\textbf{S}}\cdot(\hat{\textbf{s}}_1+\hat{\textbf{s}}_2)\nonumber
        \\&&+D\hat{S}_z^2+E(\hat{S}_x^2-\hat{S}_y^2),
\end{eqnarray}
where $\textbf{s}_1$ and $\textbf{s}_1$ refer to two radical spins.

\begin{figure}[htbp]
\begin{tabular}{c}
\includegraphics[width=7cm,height=7cm]{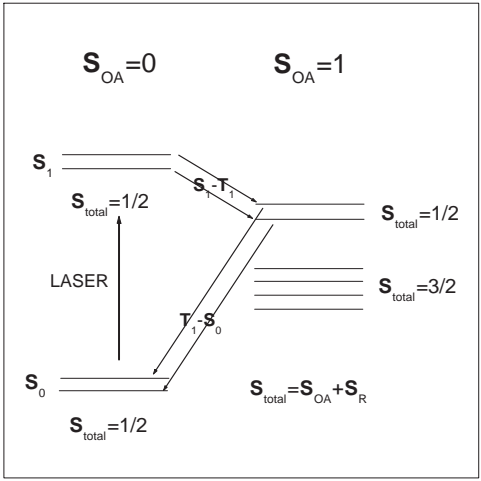}\\
(a)\\
\includegraphics[width=7cm,height=7cm]{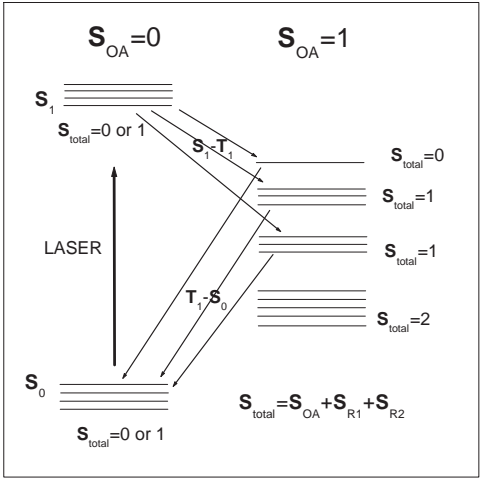}\\
(b)\\
\end{tabular}
\caption{The schematics of energy levels for different spin configurations, the photo-excitation and inter-system crossing processes. For each figure the ground and excited singlets of optical active species are shown on the left hand side while triplets on the right hand side. SRTS is shown in (a) where $\textbf{S}_{\mathrm{OA}}$ is the spin of optical active species, and $\textbf{S}_{\mathrm{R}}$ refer to the radical spin. Similar symbols apply in (b) for DRTS. }\label{pic:srtsdrts}
\end{figure}

If we further include the laser field into the Hamiltonian and adopt a rotating wave approximation (RWA), the Hamiltonian reads
\begin{eqnarray}\label{eqn:srtsh}
\hat{H}_{\mathrm{SRTS}}&=&\ket{t}[\hat{H}_{\mathrm{SRTS}}^{spin}]\bra{t}+\ket{es}[g_{r}\mu_B\textbf{B}\cdot\hat{\textbf{s}}]\bra{es}\nonumber \\&&+\ket{gs}[g_{r}\mu_B\textbf{B}\cdot\hat{\textbf{s}}]\bra{gs}\nonumber
\\&&+V(\ket{es}\bra{gs}+\ket{gs}\bra{es})
\end{eqnarray}
for single-radical-triplet system, and
\begin{eqnarray}\label{eqn:drtsh}
\hat{H}_{\mathrm{DRTS}}&=&\ket{t}[\hat{H}_{\mathrm{DRTS}}^{spin}]\bra{t}+ \ket{es}[g_{r}\mu_B\textbf{B}\cdot(\hat{\textbf{s}}_1+\hat{\textbf{s}}_2)]\bra{es}\nonumber
\\&&+\ket{gs}[g_{r}\mu_B\textbf{B}\cdot(\hat{\textbf{s}}_1+\hat{\textbf{s}}_2)]\bra{gs}\nonumber
\\&&+V(\ket{es}\bra{gs}+\ket{gs}\bra{es})
\end{eqnarray}
for double-radical-triplet system. Here $gs$, $es$, and $t$ means the ground state, excited singlet state, and triplet state, respectively. $V$ is the electrical dipole matrix element due to the laser field within RWA, and we assume that the exchange interaction is negligible when the optical active molecule is in either ground or excited singlet state \cite{wu2009}.
The dominant electrical transitional process of TR-EPR experiments for SRTS and DRTS is illustrated in Figure~(\ref{pic:srtsdrts}).
\subsection{Quantum jumping operators}
We are considering two kinds of quantum jumps operators; one is the relaxation process and the other is inter-system crossing. The relaxation processes include the relaxation of radical spins and the triplet back to the equilibrium. The electronic transitions from the excited singlet state ($S_1$) to the lowest triplet state $T_1$ will happen due to the inter-system crossing as shown in Figure~(\ref{pic:srtsdrts}), and the electronic transition from triplet down to the ground state (another inter-system crossing) is also shown in Figure~(\ref{pic:srtsdrts}) which brings the system back to the ground singlet state.

Combining the effective spin Hamiltonian and quantum jumping operators, now we can use Liuville equation to have the time evolution of the whole system under the interaction with the environment. As the Hamiltonian part of Liuvillian is just its commutator with density operator, here we list the Liuvillians describing the quantum jumps. Those Liuvillian super operators constructed by Lindblad operators are following,
\begin{eqnarray}\label{eqn:liuvs}
\hat{\hat{\mathcal{L}}}_i\rho=\sum^{n_i}_{\mu=1}{\gamma^i_{\mu}[\hat{l}^{i}_{\mu}\hat{\rho}\hat{l}^{i\dagger}_{\mu}
-\frac{1}{2}\{\hat{\rho},\hat{l}^{i\dagger}_{\mu}\hat{l}^{i}_{\mu}\}]}.
\end{eqnarray}
For \textsl{SRTS}, $i=1$ to $4$, and $\gamma^i_{\mu}$ describes how fast the quantum jumps happen.
$\hat{\hat{\mathcal{L}}}_1$ with $n_1=3$ is a super operator describing the relaxation of $\frac{1}{2}$-spin; $\hat{l}_1^1=\hat{s}^+$, $\hat{l}^1_2=\hat{s}^-$, and $\hat{l}^1_3=\hat{s}_z$, where $\hat{s}_{1z}$ operator is responsible for the pure dephasing of the quantum states.
$\hat{\hat{\mathcal{L}}}_2$ with $n_2=3$ is a super operator describing triplet spin relaxation; $\hat{l}_1^2=\hat{S}^+$, $\hat{l}^2_2=\hat{S}^-$, and $\hat{l}^2_3=\hat{S}_z$.
Here we assume that the inter-system crossing due to the spin-spin coupling is much faster than that due to the spin-orbit coupling. So the total spin angular moment should be conserved. Therefore we have the following operators for
$\hat{\hat{\mathcal{L}}}_3$ with $n_3=2$ is a super operator describing the electronic transition from excited singlet to the lowest triplet state; $\hat{l}_1^3=\ket{\frac{1}{2},\frac{1}{2}}_{t}\bra{\frac{1}{2},\frac{1}{2}}_{es}$ and $\hat{l}^3_2=\ket{\frac{1}{2},-\frac{1}{2}}_{t}\bra{\frac{1}{2},-\frac{1}{2}}_{es}$.
$\hat{\hat{\mathcal{L}}}_4$ with $n_4=2$ is a super operator describing the electronic transition from the lowest triplet to the ground state; $\hat{l}_1^4=\ket{\frac{1}{2},\frac{1}{2}}_{gs}\bra{\frac{1}{2},\frac{1}{2}}_{t}$ and $\hat{l}^4_2=\ket{\frac{1}{2},-\frac{1}{2}}_{gs}\bra{\frac{1}{2},-\frac{1}{2}}_{t}$, where spin states are written as the form $\ket{s, m_s}$.

Similarly for \textsl{DRTS}, $i=1$ to $5$ as we got another set of operators for the relaxation process of the addition radical.
$\hat{\hat{\mathcal{L}}}_1$ with $n_1=3$ is a super operator describing the relaxation of $\frac{1}{2}$-spin $\textbf{s}_1$; $\hat{l}_1^1=\hat{s}_1^+$, $\hat{l}^1_2=\hat{s}_1^-$, and $\hat{l}^1_3=\hat{s}_{1z}$.
$\hat{\hat{\mathcal{L}}}_2$ with $n_2=3$ is a super operator describing the relaxation of $\frac{1}{2}$-spin $\textbf{s}_2$; $\hat{l}_1^2=\hat{s}_2^+$, $\hat{l}^2_2=\hat{s}_2^-$, and $\hat{l}^2_3=\hat{s}_{2z}$.
$\hat{\hat{\mathcal{L}}}_3$ with $n_3=3$ is a super operator describing triplet spin relaxation; $\hat{l}_1^3=\hat{S}^+$, $\hat{l}^3_2=\hat{S}^-$, and $\hat{l}^3_3=\hat{S}_z$.
$\hat{\hat{\mathcal{L}}}_4$ with $n_4=7$ is a super operator describing the electronic transition from excited singlet to the lowest triplet state;
$\ket{1,1}^{1}_{t}\bra{1,1}_{es}$ , $\ket{1,0}^{1}_{t}\bra{1,0}_{es}$, $\ket{1,-1}^{1}_{t}\bra{1,-1}_{es}$, $\ket{1,1}^{2}_{t}\bra{1,1}_{es}$ , $\ket{1,0}^{2}_{t}\bra{1,0}_{es}$, $\ket{1,-1}^{2}_{t}\bra{1,-1}_{es}$, and $\ket{0,0}_{t}\bra{0,0}_{es}$, where  we use superscripts $1,2$ as there are two triplet manifolds when $S_{\mathrm{OA}}=1$; one is from $1\oplus1$ and another from $1\oplus0$.
$\hat{\hat{\mathcal{L}}}_5$ with $n_5=7$ is a super operator describing the electronic transition from the lowest triplet to the ground state; $\ket{1,1}_{gs}\bra{1,1}^1_{t}$ , $\ket{1,0}_{gs}\bra{1,0}^1_{t}$, $\ket{1,-1}_{gs}\bra{1,-1}^1_{t}$, $\ket{1,1}_{gs}\bra{1,1}^2_{t}$ , $\ket{1,0}_{gs}\bra{1,0}^2_{t}$, $\ket{1,-1}_{gs}\bra{1,-1}^2_{t}$, and $\ket{0,0}_{gs}\bra{0,0}_{t}$.

\section{Numerical simulations for EPR experiments of SRTS and DRTS}
By now we have Hamiltonians including equation~(\ref{eqn:srtsh}) and equation~(\ref{eqn:drtsh}) for the coherent process driven by zero-field splitting, exchange interaction, magnetic field, and laser field. In addition, the dissipative process of the whole system is described by the Livillians defined in equation~(\ref{eqn:liuvs}). So next we are going to put in some reasonable parameters and calculate the time evolution of the density operator, especially the populations for the excited singlet state and triplet state and the field dependency of EPR spectra due to a perturbation of transverse microwave field, i.e., a linear approximation for the EPR spectra. In general, the relaxation of triplet is much faster than that of radical spins by about hundred times \cite{yamauchi2004} at liquid Nitrogen temperature, the inter-system crossing is comparable with the relaxation of triplet, and the decay from the triplet to the ground singlet state is very slow which is about $ms$. In the following we will vary these parameters and take a close look at the effect of these relaxation processes. Another important parameter is the exchange coupling between radical and triplet which can be both calculated by density functional theory \cite{wu2009} and estimated through the broadening of EPR spectra which is set to be $-10 \ \mathrm{mK}$. \textsl{In the following we will use $\mathrm{mK}$ as the energy unit}.  So in this section, we will first concentrate on SRTS and DRTS respectively, where we can vary the different relaxation parameters as our model could predict the time evolution of density operator and EPR spectra with any parameters in principle; meanwhile we present the 3-dimensional figure for the time-resolved EPR spectra. And then we will study the correlations between radical spins in DRTS caused by the random magnetic field driven by the relaxation of triplet.

\subsection{Time evolution of density operator}
Zero-field splitting $D$ and $E$ can be measured from the precursors; $D$ is about $20 \mathrm{mK}$ and $E$ $3 \mathrm{mK}$. As we have $12$ parameters left for MonoTEMPO, in order to keep our model as simple as possible, for the moment we assume that $g_r=g_t=2.0$ because the observation of the interaction between radical spin and triplet is somewhat independent of spin-orbital coupling which is strongly related to g-factor, and we also reck on that $\gamma^1_1=\gamma^1_2, \gamma^1_3=0$, $\gamma^2_1=\gamma^2_2, \gamma^2_3=0$ as for high temperature this is almost true, $\gamma^3_1=\gamma^3_2$, and $\gamma^4_1=\gamma^4_2$ i.e., the inter-system crossing rate and the decay rate are spin-independent for the moment. Having some simplifications for these parameters, we are going to use some reason values captured by the previous literature for the relaxation processes and inter-system crossing, and our calculated exchange interaction to compute the spectra to illustrate the physics inside the system, and then we can compare it with the experiments, so we can have some physical insights from these comparisons. From these simulations, we can see that the EPR spectra at high temperature can be predicted by using our programs once if we have know the details of the relaxation and inter-system crossing. We set the parameters as: $J\simeq -10 \mathrm{mK}$, $\gamma^1_1=0.067 \mathrm{mK}$, $\gamma^2_1=67.0 \mathrm{mK}$, $\gamma^4_1=0.035 \mathrm{mK}$, and $V$ is about $0.67 \mathrm{mK}$. The magnetic field is large compared with the above parameters which is about $200 \mathrm{mK}$. Now we can first compute time evolution of density operator withe different inter-system crossing rates.

\begin{figure}[htbp]
\begin{tabular}{c}
\includegraphics[width=8.5cm,height=6.5cm]{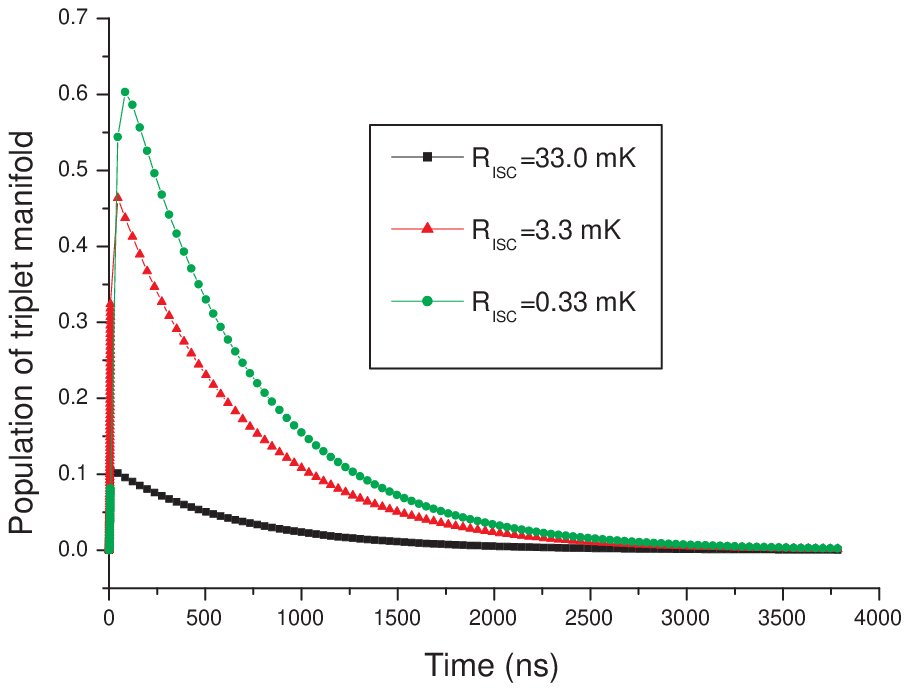}\\
(a)\\
\includegraphics[width=8.5cm,height=6.5cm]{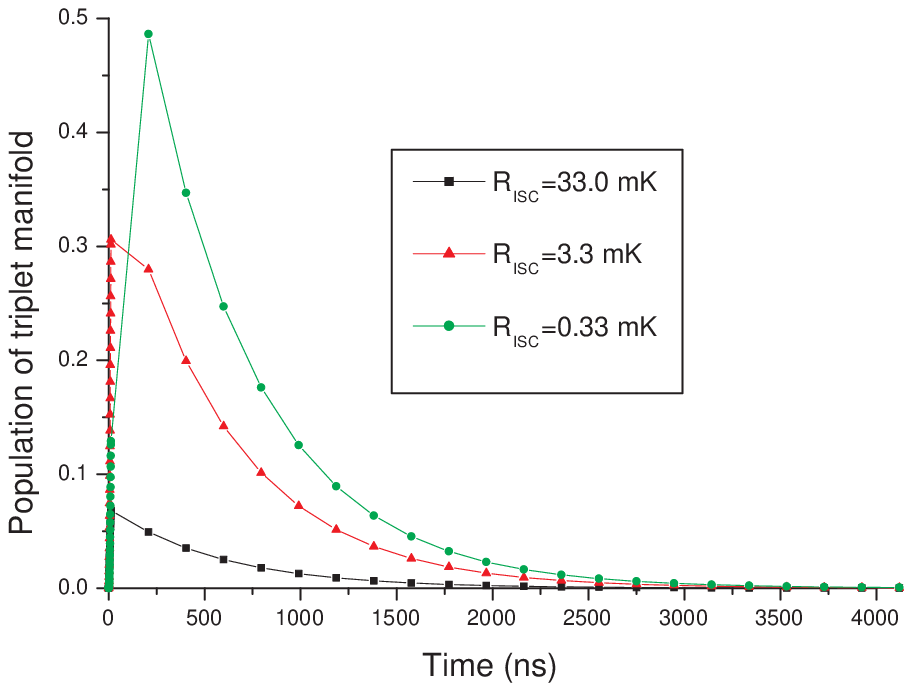}\\
(b)\\
\end{tabular}
\caption{(Color online.) The population of the manifold with $\textbf{S}_{\mathrm{OA}}=1$ as a function of time with varying inter-system crossing rate $R_{\mathrm{ISC}}=\gamma^3_1$ for SRTS and $R_{\mathrm{ISC}}=\gamma^4_1$ for DRTS is shown for both (a) SRTS and (b) DRTS. Notice that as $R_{ISC}$ increases from $0.33 \ \mathrm{mK}$ (black curve with square symbols), via $3.3 \ \mathrm{mK}$ (red curve with triangle symbols) to $33 \ \mathrm{mK}$ (green curve with circle symbols), the population of triplet becomes less during laser pumping for SRTS and DRTS. The slight difference between them is because there are more channels to dump the populations in triplet into singlet ground state.}\label{pic:monocompleterhodt}
\end{figure}

\textsl{Inter-system crossing rate dependency.}
In this part, we simulate the case where the laser is turned on from time $0$, then turned off completely at $t_1\simeq 8.0 \ ns$, and the density operator is evolved till the time $t_2\simeq 4000 \ ns$. We first plot the triplet population as a function of inter-system crossing rate $\gamma^3_1$ as shown in Figure~(\ref{pic:monocompleterhodt}) because the population of triplet manifold is immediately related to the strength of EPR signals. We can see that the population of triplet manifold will be higher when we tuned the inter-system crossing rate to be smaller. This is because when the population transfers from singlet excited state to triplet, the zero-field splitting and the fast relaxation of the triplet will transfer the population into $S_{total}=2$ state which will go back down to the ground much more slowly. So the more slowly the population is transferred, the more population will be left in the triplet manifold.

\begin{figure}[htbp]
\begin{tabular}{c}
\includegraphics[width=8.5cm,height=6.5cm]{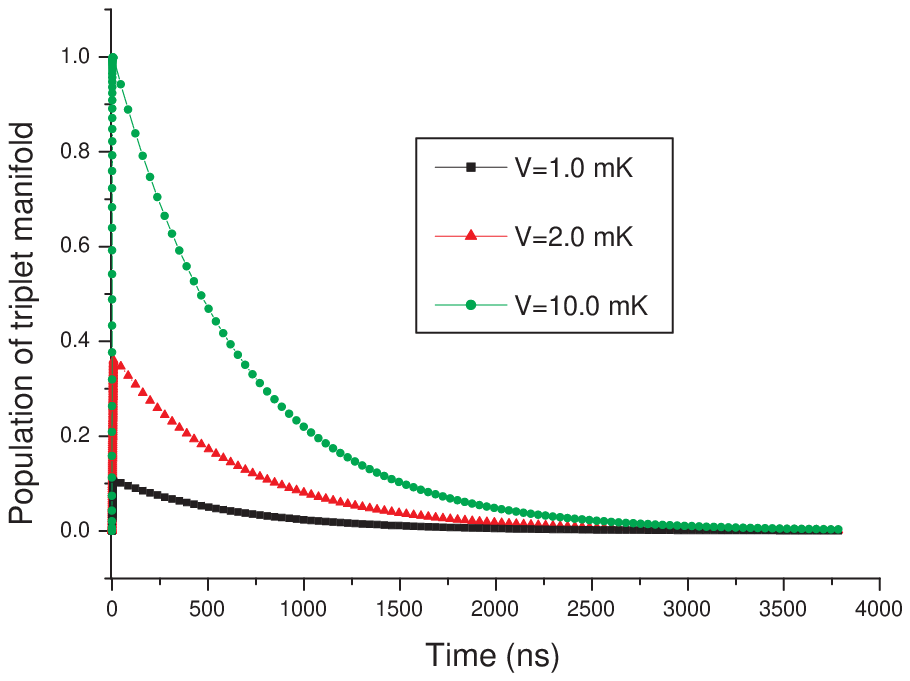}\\
(a)\\
\includegraphics[width=8.5cm,height=6.5cm]{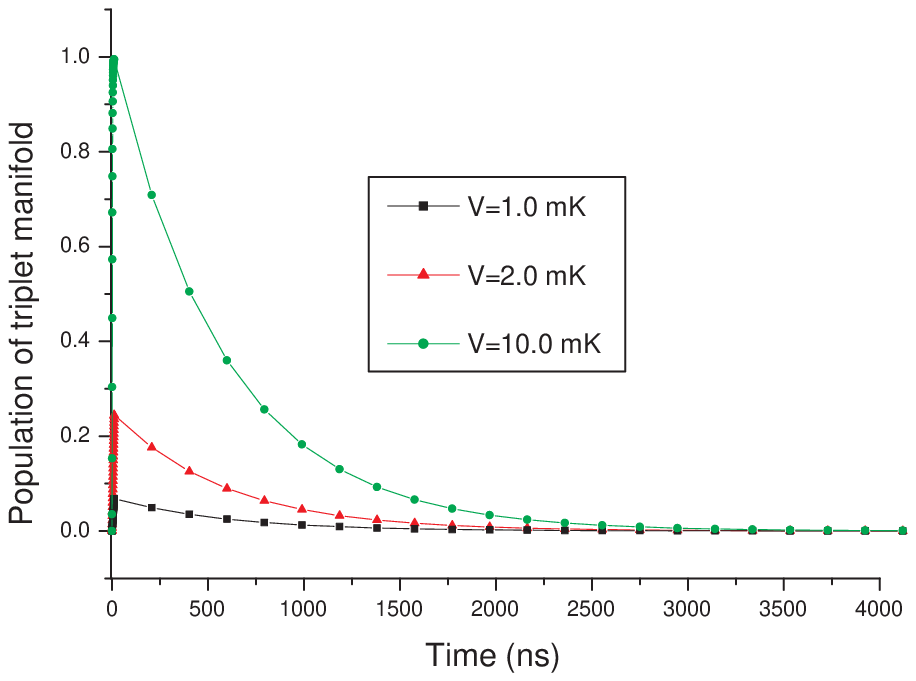}\\
(b)\\
\end{tabular}
\caption{(Color online.) The population of the manifold with $\textbf{S}_{\mathrm{OA}}=1$ as a function of time with different values of $V$ is shown for both (a) SRTS and (b) DRTS; black curve with square symbols corresponds to $V=1.0 \ \mathrm{mK}$, red curve with triangle symbols $V=2.0 \ \mathrm{mK}$, and green curve with circle symbols $V=10.0 \ \mathrm{mK}$. Notice that as $V$ increases, i.e., the laser pumping is strengthened, the population of triplet becomes richer.}\label{pic:monocompleterhov}
\end{figure}

\textsl{Laser-field strength dependency.}
As shown in Figure~(\ref{pic:monocompleterhov}), we fix the inter-system crossing rate to be $33 \mathrm{mK}$, and we see that as optical pumping is strengthened, there will be more population in the triplet manifold.

\textsl{Decay rate dependency.}
We can also show that the population of triplet manifold will drop faster when decay rate becomes larger as expected. The population of triplet manifold as a function of time with different decay rates is shown in Figure~(\ref{pic:monocompleterhov}).
\begin{figure}[htbp]
\begin{tabular}{c}
\includegraphics[width=8.5cm,height=6.5cm]{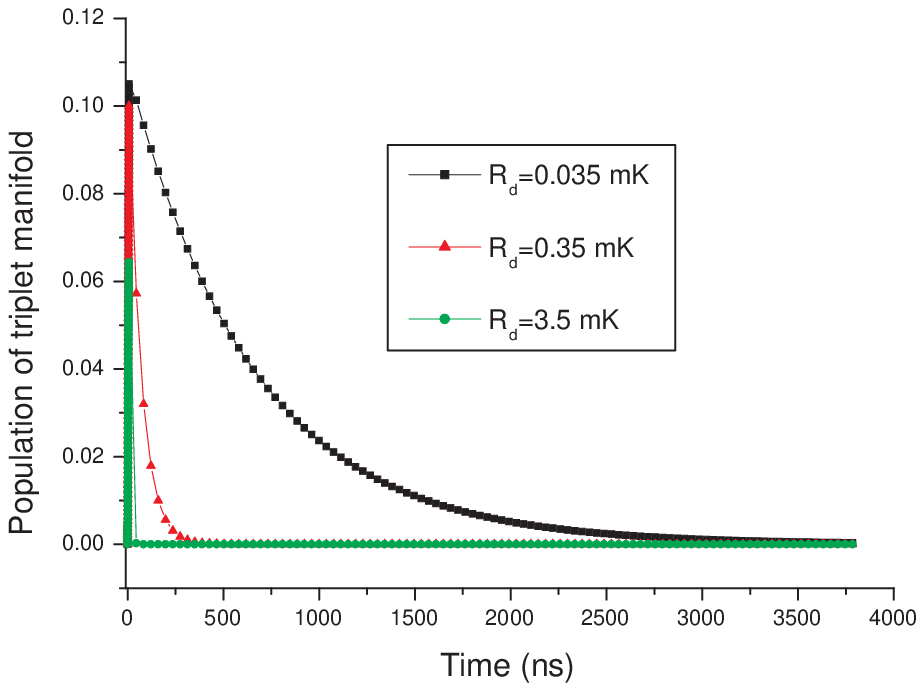}\\
(a)\\
\includegraphics[width=8.5cm,height=6.5cm]{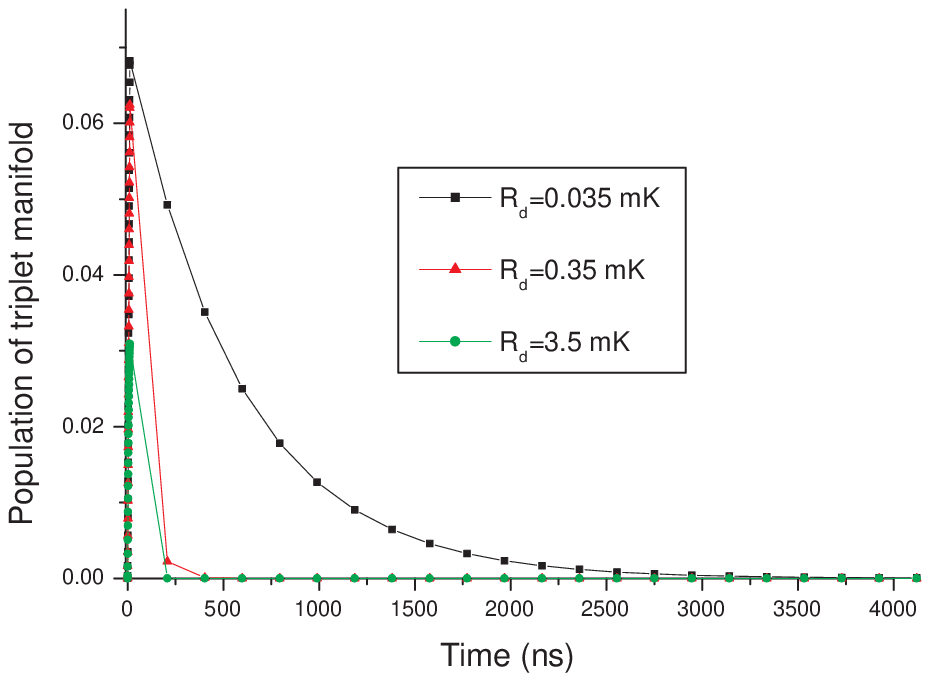}\\
(b)\\
\end{tabular}
\caption{(Color online.) The population of the spin manifold with $\textbf{S}_{\mathrm{OA}}=1$ as a function of time with different values of decay rates $R_d=\gamma^4_1$ for SRTS and $R_d=\gamma^5_1$ for DRTS is shown for both (a) SRTS and (b) DRTS; black curve with square symbols corresponds to $=1.0 \mathrm{mK}$, red curve with triangle symbols $V=2.0 \mathrm{mK}$, and green curve with circle symbols $V=10.0 \mathrm{mK}$. Notice that as $R_d$ increases, the population of triplet decay faster.}\label{pic:rhotd}
\end{figure}

\subsection{EPR spectra}
As discussed in \S\ref{sec:1}, we EPR spectra should be proportional to the response function of an open quantum system stated in equation~\ref{eqn:chinonstationary}. From equation~\ref{eqn:chinonstationary}, we can also see that EPR spectra is dependent both on time and magnetic field which results from the fact that the density operator is non-stationary for the unperturbed Hamiltonian. To simulate EPR spectra we take the perturbation $\mathcal{B}$ as the commutator $[\hat{S}_x+\hat{s}_x,\cdot]$ for SRTS and $[\hat{S}_x+\hat{s}_{1x}+\hat{s}_{2x},\cdot]$ for DRTS, and the observable to be $\hat{S}_x+\hat{s}_x$ for SRTS and $\hat{S}_x+\hat{s}_{1x}+\hat{s}_{2x}$ for DRTS. Essentially, the physical meaning is that we put in a perturbation which flips the spins and then after some time we see the response of the system under the interaction with the environments. EPR spectra is Fourier transformation of the time-dependent response to the energy domain.

\begin{figure}[htbp]
\begin{tabular}{c}
\includegraphics[width=8.5cm,height=6.5cm]{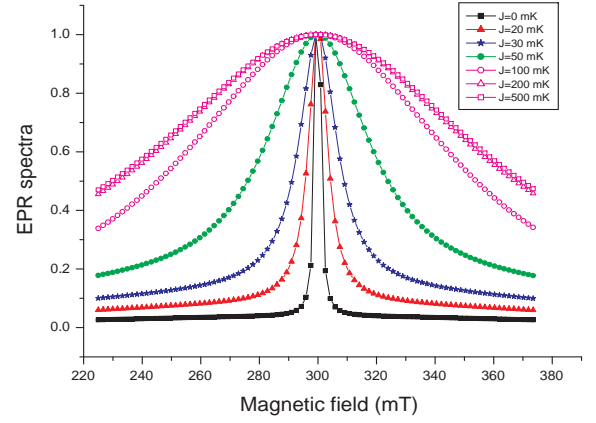}\\
(a)\\
\includegraphics[width=8.5cm,height=6.5cm]{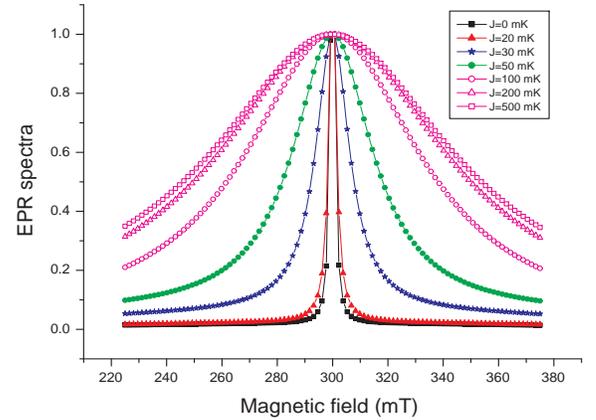}\\
(b)\\
\end{tabular}
\caption{(Color online.) The normalized EPR spectra as a function of magnetic field for both (a) SRTS and (b) DRTS are shown. The exchange interaction $J$ is varied as taking values $0$ (black square), $20$ (red triangle), $30$ (blue star), $50$ (green circle), $100$ (magenta hollow circle), $200$ (magenta hollow triangle) and $500 \ \mathrm{mK}$ (magenta hollow square). Notice spectra broadening and the width saturation. }\label{pic:monoeprofj}
\end{figure}

\textsl{Exchange-interaction-dependency.}
The main purpose that we use EPR for radical-triplet system is to look at how strong the exchange interaction between radical and triplet is. From Figure~(\ref{pic:monoeprofj}) we see that the broadening of EPR spectra is strongly dependent on the strength of exchange interaction $J$ which is crucial both for quantum gate operations in QIP and molecular magnetic switching in spintronics. In Figure~(\ref{pic:monoeprofj}) we keep all the other parameters the same as in Figure~\ref{pic:monocompleterhodt} with $R_{ISC}=33 \mathrm{mK}$, $V=1.0$, and $t_0=1.0$ except varying $J$ to $0 \mathrm{mK}$, $20 \mathrm{mK}$, $30 \mathrm{mK}$, $50 \mathrm{mK}$, $100 \mathrm{mK}$, $200 \mathrm{mK}$ and $500 \mathrm{mK}$. From this Figure, we can see that the line shape broadening of EPR spectra is strongly dependent on the strength of exchange interaction; as the exchange interaction becomes stronger, the line shape will be broader. When $J=0 \ \mathrm{mK}$, the spectra is dominated by the perturbing oscillating field frequency. When $J\neq 0$, it will spread the eigenvalues of the effective spin Hamiltonian, so we see the broadening. And when $J\sim 500 \mathrm{mK}$, the broadening will be saturated as the broadening is constrained by the spin-lattice relaxation time of triplet.

\begin{figure}[htbp]
\begin{tabular}{c}
\includegraphics[width=8.5cm,height=6.5cm]{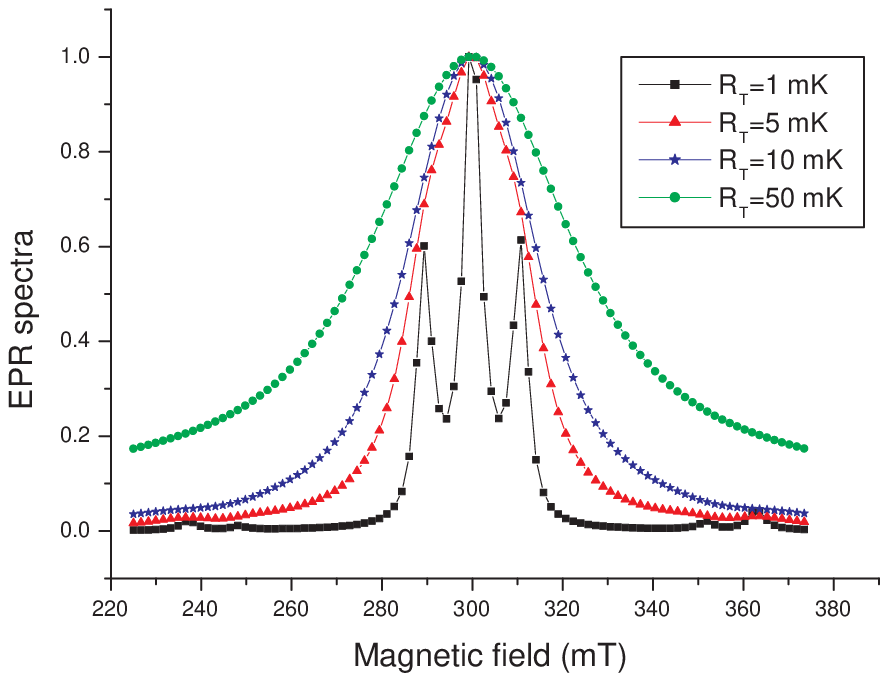}\\
(a)\\
\includegraphics[width=8.5cm,height=6.5cm]{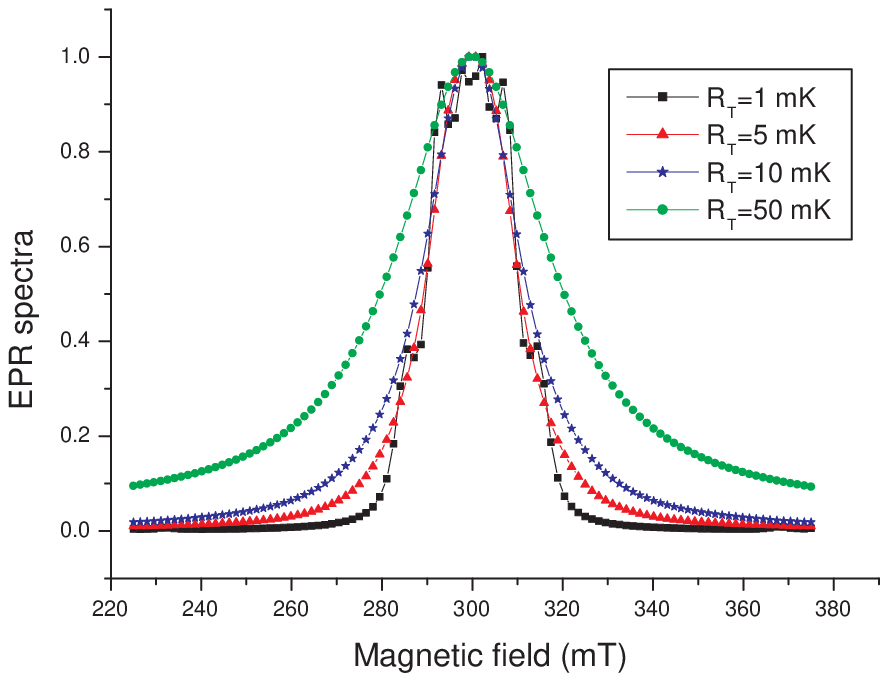}\\
(b)\\
\end{tabular}
\caption{(Color online.) The normalized EPR spectra of SRTS and DRTS as a function of magnetic field with varied triplet relaxation rate $R_t$ taking values $1$ (black square), $5$ (red triangle), $10$ (blue star), $50 \ \mathrm{mK}$ (green circle). Notice spectra splitting due to the long-living triplet and its disappearance as the relaxation of triplet becomes faster. }\label{pic:monoeprofrt}
\end{figure}

\textsl{Triplet relaxation-rate-dependency}.
In the previous calculations we always assume that the relaxation of triplet back to equilibrium is very fast. It is interesting to see what happens if we change the triplet relaxation rate. Here we keep all the parameter the same as in Figure~(\ref{pic:monoeprofj}) except $J= -50 \ \mathrm{mK}$ and the triplet relaxation rate $R_t=\gamma^2_1$ taking the values $1.0 \ \mathrm{mK}$, $10.0 \ \mathrm{mK}$, $50.0 \ \mathrm{mK}$, and $100 \ \mathrm{mK}$. As we see Figure~(\ref{pic:monoeprofrt}), the relaxation rate of triplet is also responsible for the broadening of EPR spectra. And the more important is that EPR spectra will be changed significantly if the triplet relaxation is as slow as the relaxation of radical spin, i.e., $\gamma^1_1\approx \gamma^2_1$. When $R_T=1 \ \mathrm{mK}$, the typical EPR spectra of a triplet with non-zero field splitting will appear. Meanwhile, as the relaxation rate becomes larger, the triplet feature will be gone; we can only see the radical $\frac{1}{2}$-spin.

\textsl{Analysis of EPR spectra by decomposing to the different spin correlations}
\begin{figure}[htbp]
\begin{tabular}{c}
\includegraphics[width=8.5cm,height=6.5cm]{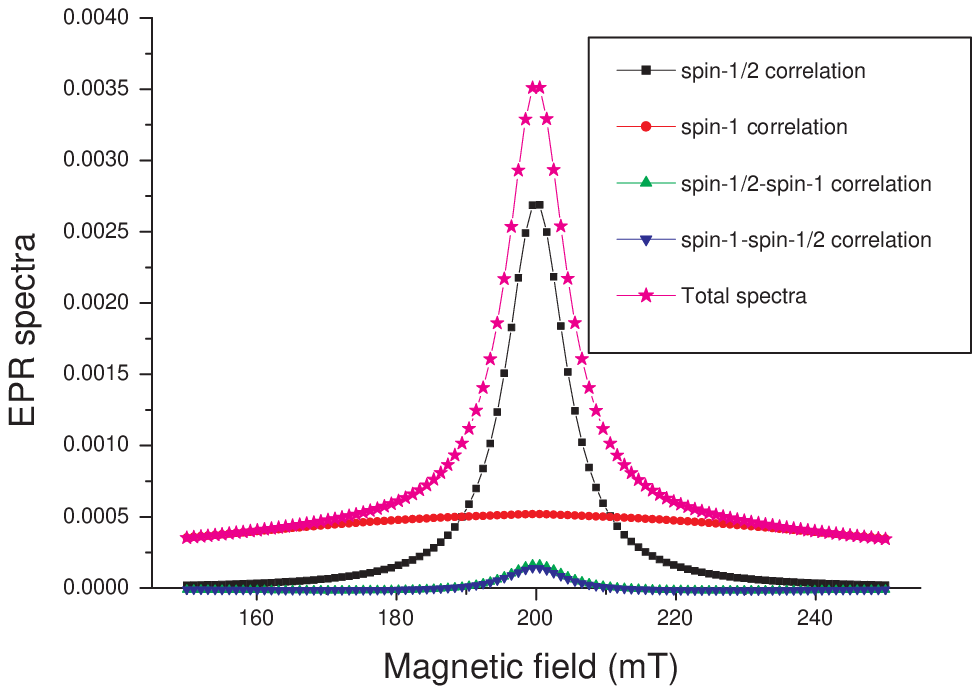}\\
(a)\\
\includegraphics[width=8.5cm,height=6.5cm]{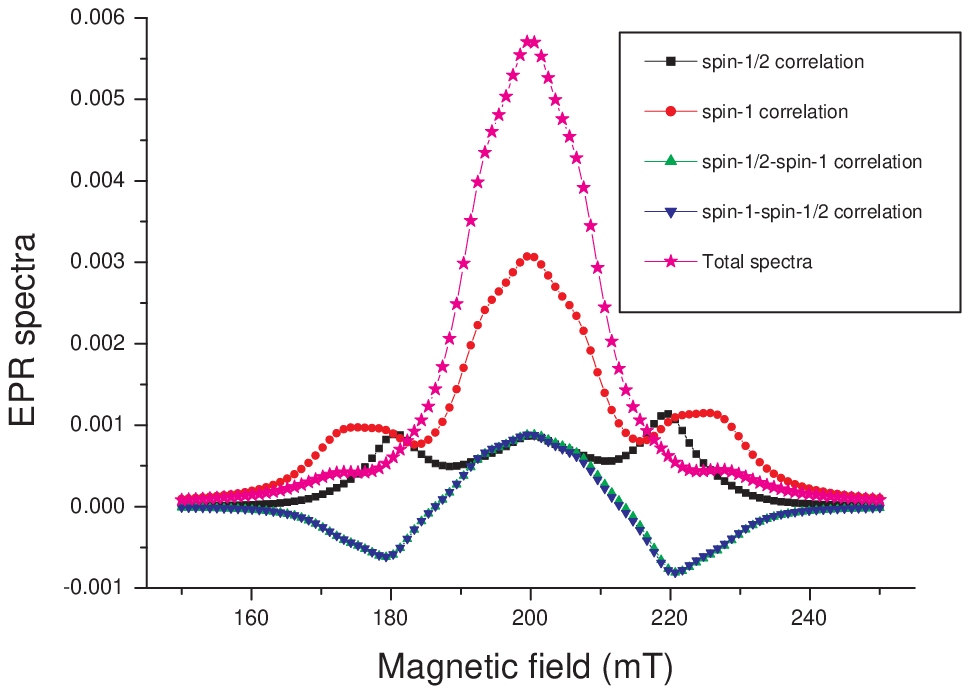}\\
(b)\\
\end{tabular}
\caption{(Color online.) The EPR spectra of SRTS as a function of magnetic field with varied triplet relaxation rate $R_t$ taking values $67 \ \mathrm{mK}$ (a) and (b) $0.67 \ \mathrm{mK}$ while other parameters are kept the same as in Figure~(\ref{pic:monoeprofrt}).}\label{pic:monoeprofrtdecompose}
\end{figure}

\textsl{Time-dependent EPR spectra}. The time-resolved EPR spectra is crucial for observations of the interactions between radical and triplet. Here we present the 3-dimensional picture for the calculated EPR spectra as a function of time and magnetic field.
\begin{figure}[htbp]
\begin{tabular}{c}
\includegraphics[width=8.5cm,height=6.5cm]{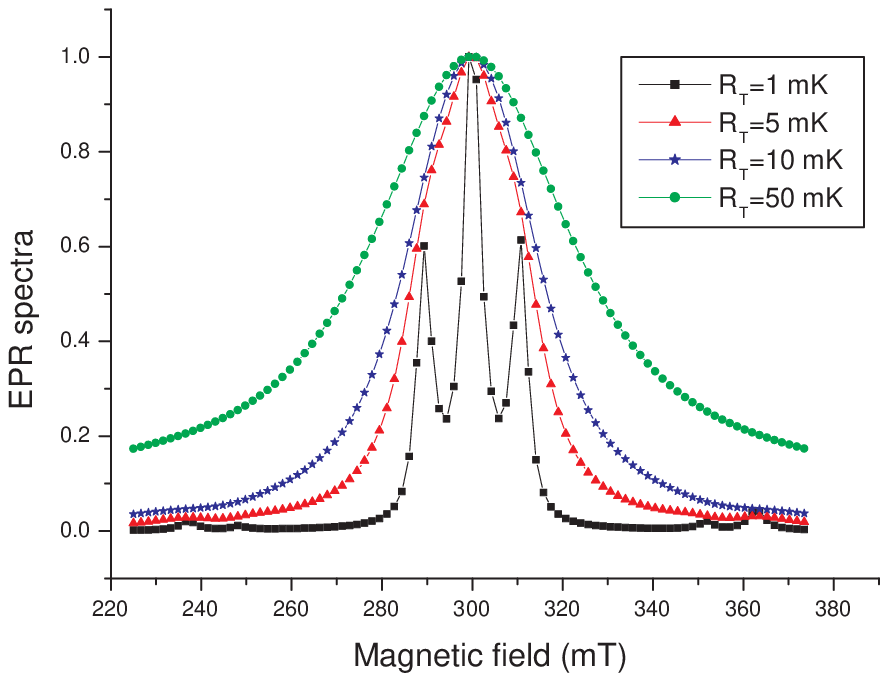}\\
(a)\\
\includegraphics[width=8.5cm,height=6.5cm]{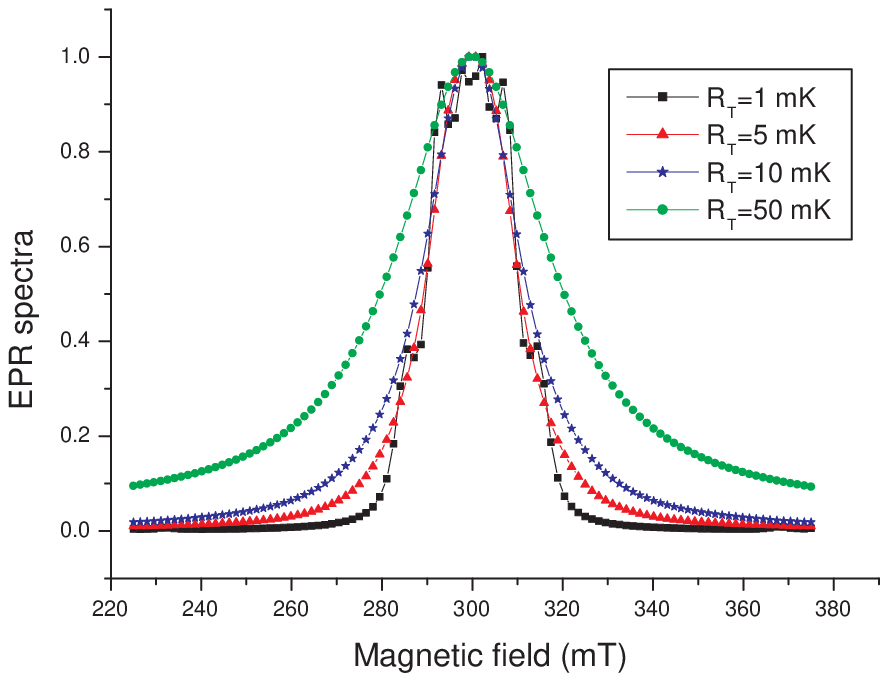}\\
(b)\\
\end{tabular}
\caption{(Color online.) The EPR spectra of SRTS and DRTS as a function of magnetic field with varied triplet relaxation rate $R_t$ taking values $1$ (black square), $5$ (red triangle), $10$ (blue star), $50 \ \mathrm{mK}$ (green circle). Notice spectra splitting due to the long-living triplet and its disappearance as the relaxation of triplet becomes faster. }\label{pic:monoeprofrt3d}
\end{figure}

\section{Conclusion and discussions}

\end{document}